# Optical and transport properties of low-dimensional semiconductor nanostructures

K. Král[1,*], M. Menšík[2]

[1]Institute of Physics, ASCR, v.v.i., Na Slovance 2, 18221 Prague 8, Czech Republic
[2]Institute of Macromolecular Chemistry, ASCR, v.v.i., Heyrovského nám. 2, 162 06 Praha 6, Czech Republic

*e-mail: kral@fzu.cz

**Abstract.** The interpretation of the electronic kinetic processes in the quantum zero dimensional nanostructures is considered. The main mechanism of the processes is supposed to be the interaction of electrons with the optical phonons. An emphasis is put on the recently measured effect of the long-time photoluminescence of quantum dot samples, which is observed to occur after an illumination of the sample by a laser pulse. In addition to this, an attention is devoted to the possible origin of the optical effect of the blinking (intermittence) of the optical emission of certain quantum dot samples under a permanent optical excitation, and to another similar effect.

### 1. Introduction

Small quantum semiconductor nanosystems display sometimes properties, which are qualitatively different from the bulk samples [1]. The electronic transport properties are usually dependent on the character of the scattering of the charge carriers in the nanostructure. In a low-dimensional semiconductor nanostructure the scattering electron has restricted possibilities as far as the dimensionality of the space in which the electron can leave the scattering target after a collision is concerned. In the extreme case of a quantum dot, or a zero-dimensional nanoparticle, the electron cannot leave the scattering target after a collision. It reflects at the boundary of the quantum dot and returns back to the target, to continue the quantum mechanical act of scattering. The scattering target can be for example the modes of the optical phonons of the atomic lattice vibrations of the quantum dot. One can say that in the zero-dimensional nanostructures there are natural conditions for taking into account the perturbations, influencing the electronic motion, in a multiple scattering approximations. This could be so even for such scattering mechanisms of electron in a zero-dimensional nanostructure which in the case of a bulk system is regarded as weak.

The conditions for the use of the multiple scattering of an electron on the scattering center may lead to a necessity to use a suitable tool for treating the infinite series of the expansion of the calculated result in the powers of the perturbation of the electronic motion. In practice it means that the Golden rule formula, the standard theoretical equipment of a solid state physicist of today, used often in the bulk systems with an advantage, may be not enough suitable for achieving the expected goal of the theoretical effort in the case of the zero-dimensional system. From this reason, the theoretical approach to the electronic transport may need such techniques like the nonequilibrium Green's functions [2].

The multiple scattering of electrons in a zero-dimensional nanostructure is at the present time often not taken into account [3] and for example, the rapidity of the electron energy relaxation in quantum dots is explained by the Auger mechanism of the electron energy relaxation [4]. Nevertheless, the earlier considered existence of the optical phonon bottleneck in the quantum dots, expected to exist in quantum dots on the grounds of the perturbation calculation argumentation, remains so far not yet verified [5].

In the present work the electronic transport in the zero-dimensional nanostructures will be treated with the help of the nonequilibrium Green's functions and with the quantum kinetic equations for the description of the irreversible processes in the electronic motion in the environment of the atomic vibrations.

The significance of our understanding the electronic properties of the zero-dimensional nanostructures is unquestionable. It may help to comprehend the problems of quantum computation and quantum information processing in the future hardware. We may also progress towards the understanding of properties and processes in the molecule DNA [6] and in other basic elements of the living matter [7].

We shall point out some the main characteristics of the electron kinetics in quantum dots in the presence of the electron interaction with the optical phonons. The attention will be turned to the time development of the decay of the photoluminescence of the quantum dot samples after the illumination by a laser pulse. We shall show the possible origin of the power law dependence of the photoluminescence intensity. We shall then pay the attention to another effect, which also displays power law time characteristics, namely to the effect of the blinking of quantum dots, in other words to the intermittence of the optical emission of the quantum dot sample under a steady state illumination by excitation by light.

### 2. Electron energy relaxation in quantum dots

We shall summarize some of the main properties of the electron energy relaxation mechanism based on the interaction of electrons with the optical phonons. This mechanism be effective even in the limiting case of very low density of excited electrons in the quantum dot sample.

After an illumination by light electrons and hole pairs can be excited into the above barrier states of the environment of a quantum dot. Finally electrons and holes can be captured by quantum dots. Because of the electrostatic forces between electrons and holes an electron coupled to a hole is supposed to be captured by single quantum dot. Neglecting for simplicity the electron-hole electrostatic coupling we can consider the hole motion as independent from that of the electron. The hole particles are usually heavier than the electrons in the conduction band states of the quantum dot, so that their bound states in the dot are more dense on the energy scale. From this reason it is often assumed that the holes relax quickly to their ground state in the dot via the low-order perturbation calculation processes. The electron energy relaxation is then treated as a nonequilibrium process with using the quantum kinetic equations [8-12].

The confinement of the electron motion in all three dimensions results in the use of the self-consistent Born approximation to the electronic self-energy for the inclusion of the interaction of the electron with the optical phonons. In the diagonal approximation the relaxation rate formula has the form:

$$\frac{dN_1}{dt} = -\frac{2\pi}{\hbar}\alpha_{01}$$

$$\times \left[ N_1(1-N_0)\left( (1+\upsilon_{LO})\int_{-\infty}^{\infty} dE \sigma_1(E)\sigma_0(E-E_{LO}) \right.\right.$$

$$\left. +\upsilon_{LO}\int_{-\infty}^{\infty} dE \sigma_1(E)\sigma_0(E+E_{LO}) \right) \quad (1)$$

$$-N_0(1-N_1)\left( (1+\upsilon_{LO})\int_{-\infty}^{\infty} dE \sigma_0(E)\sigma_1(E-E_{LO}) \right.$$

$$\left.\left. +\upsilon_{LO}\int_{-\infty}^{\infty} dE \sigma_0(E)\sigma_1(E+E_{LO}) \right) \right].$$

In this formula we assume that the quantum dot has two electronic states for which as the wave functions we take the ground state ($n=0$) and one of the lowest energy excited states ($n=1$) of the infinitly deep cubic quantum dot with effective mass of InAs. The material parameters in the electron-phonon interaction constant $\alpha_{01}$ are those of InAs. The quantity $N_n$ is the occupation of the $n$-th electronic state. The quantity $\sigma_n$ is electronic spectral density of the $n$-th state. The later formula, giving the relaxation rate of the electron from the electronic excited state 1 to the electronic ground state 0, due to the electron interaction with the longitudinal optical phonons in the quantum dot, see refs. [8-12], does not contain the energy conservation delta function of energy variable. This is an implication of the self-consistent Born approximation to the self-energy. We discussed the meaning of the present approximation in earlier publications [8-11].

One of the significant properties of the present approximation to the electron self-energy is the upconversion of the occupation of the electronic energy levels (see e.g. [11]). This effect can be explained as an implication of the multiple ecattering of the electron on the optical phonons, which leads to virtual coherent multiphonon states. These states of the bosonic modes can be understood as corresponding to classical microscopic vibrations of the atomic lattice. The presence of such lattice oscillations can make the effective Hamiltonian for the electron as time dependent, and the system non-conservative. The two important subsystems, the electron, and the optical phonon system, do not then exchange only heat. We therefore cannot expect that they will behave as two sybsystems having the same effective temperature. This property is an implication of the smallness [1] of the quantum dot system and should not be unexpected to be met also in other small systems. A number of experimentally found properties of quantum dot properties have recently been discussed from the point of view of the upconversion of the electronic energy level occupation [13,14]. We shall pay a more detailed attention to a manifestation of the upconversion effect in the next section.

### 3. Power-law decay of quantum dot photoluminescence

The time dependence of the decay of the photoluminescence after an illumination by a laser pulse may be significant from the point of view of applications of the zero-dimensional semiconductor nanostructures. In recent experiments it has been demonstrated that Stranski-Krastanow method grown quantum dot samples can display a time dependence of the decay of the luminescence that reminds the power-law particle decay discussed earlier in connection with mesons [15,16].

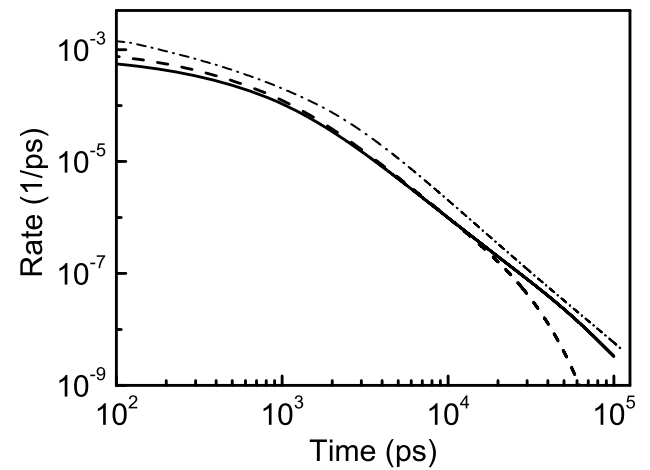

*Fig.1 The rate of the upconversion* $r = \frac{dN_1}{dt}$ *for cubic dot of InAs with the lateral size of 15 or 12 nm. $N_1$ is kept equal zero throughout the evolution in time. The full and dashed lines correspond to the dot with the lateral size of 15 nm. The full line corresponds to both the temperatures 10 K and 30 K, while the dashed line is computed at 50 K. The dash-dot line is computed for the lateral size of the dot equal to 12 nm at the lattice temperature of 10 K. In the dot with the size of 15 nm* $(E_1 - E_0)/E_{LO} \cong 7.2$, *for which case the interlevel separation* $E_1 - E_0$ *is about 217 meV in this model of quantum dot. In the quantum dot with the lateral size of 12 nm the ratio of* $(E_1 - E_0)/E_{LO}$ *equals about 11.3, which gives about 341 meV for the electronic interlevel separation.*

We remind the power-law decay of the photoluminescence signal as it was detected by T. Shamirzaev *et al.* in papers [17,18]. The authors of the experiments [17,18] suggest to explain the power law decay to be due to electron-hole pair being blocked in the dark state of the ground state of the triplet exciton. They assume that at low temperatures and in small quantum dots the large exchange interation between these two particles makes the thermal escape of the electron to a singlet state not probable. A set of inter-dot tunneling channels is then expected to participate on the observed power-law time dependence of the measured photoluminiscence intensity.

In our model, we suggest to explain the power-law time dependence as due to the electron-phonon interaction in the quantum dot. In analogy to the papers [17,18] we also assume that the basic factor in the effect of the slowing-down of the photoluminescence is the electron captured in the triplet exciton state in the quantum dot. However, instead of assuming the existence of a set of interdot tunneling mechanisms, which should open the way to the electron-hole annihilation, we utilize the interaction of the electron with the optical phonon modes in the quantum dot, taken into account in the self-consistent Born approximation to the electronic self-energy [19].

In our model calculation of the photoluminescence time dependence we assume that at the beginning of time the electron is placed in its ground state of the triplet exciton made with the hole particle, so that the optical emission channel is not allowed. We assume that in the small quantum dots there are only two orbital bound states of the electron. It is assumed that once the electron reaches the excited state, it can immediately leave the quantum dot and be transferred to another quantum dot in which it meets a hole with a suitable spin, to recombine with, by emitting light.

So that the main mechanism responsible for the power law time dependence is then the upconversion mechanism, which delivers the electronic density from the ground state to the excited state. It was shown some time ago that with decreasing the electronic occupation of a sigle dot, the effectiveness of the multiple phonon effect of the electron-phonon upconversion mechanism decreases [20]. This could be the reason for the slowing down the luminescence intensity decay to the power-law time dependence. We calculate numerically the time derivative of occupation of the quantum dot and regard this quantity as proportional to the observed intensity of the photoluminescence signal after the preparation of the electron in the quantum dot by the laser pulse. In the Fig. 1 it is shown that at long times the time dependence of the photoluminescence signal (dash-dot line) really comes out as a power-law functional dependence on the time variable.

In our model using the electron-phonon interation we circumvented the assumption of an existence of a number of inter-dot tunneling mechanism, however our assumption of the electron's ability to be quickly transferred from the excited state to another dot with a hole in a state suitable for the light emission means that this part of the model deserves a further attention. The significance of the present model is in that it shows an ability of the electron-phonon interaction to give the power law dependence of the signal, in the case when the nonadiabaticity of the electron-phonon interaction is taken into account. Besides other examples of agreement with experimental data, of the theoretical conclusions achieved with using the self-consistent Born approximation to the electronic self-energy, this model explanation of the power-law decay serves a reason to expect that even some other properties of the quantum dots may be influenced to some extent by the electron-phonon interaction. In the next section we present a brief report about our arguments given in support of the significant role of the electron-phonon interaction in the so called blinking (intermittence) effect in the light emission of the quasi-zero-dimensional nanostructures [21].

## 4. Blinking of quantum dots

Especially the colloidal quantum nanoparticles (quantum dots) display the effect of the intermittence of the light emission of the quantum dot sample under a steady state excitation by a light source [21]. On the other hand the Stranski-Krastanow grown samples of the self-assembled quantum dots usually do not display this effect. In the Stranski-Krastanow dots the quantum dots themselves are grown in such a way that they remain in an atomic contact with the substrate. This contact could provide an efficient channel for the cooling of the quantum dot when the dot is being heated by some process, say by electron energy relaxation effect, when a captured electron relaxes down to the ground state within a dot.

Although a suitable detailed balance of the heat transfer in the Stranski-Krastanow and in the colloidal quantum dots is probably not yet available, we might have an opinion that in the colloidal quantum dots the vibrational optical modes receive certain amount of energy from the relaxing electrons and this energy leads to certain manifestations of an energy being accumulated in these modes. The phonon modes gaining energy from the energy relaxing electrons can become excited into coherent phonon states, which could be for simplicity represented by classical vibrations of the oscillator modes of the optical phonons. We expect the excitation of the particularly coherent phonon modes, because the self-consistent Born approximation, which could be considered as successfully interpreting the process of the relaxation of the electron energy in quantum dots,

The classical oscillators of the optical phonon vibrations can in such a way obtain an amount of energy of the oscillatory motion so that they reach the regime in which the nonlinearities of their scillations become significant [22]. Especially, after a period of time, during which the captured electrons deliver an amount of energy into a given vibrational mode, this mode can start to display a chaotic motion.

Let us express now a working hypothesis about a role played by the chaotic state of motion of the phonon mode which otherwise plays a medium through which a captured electron relaxes energy when attempting to get to the ground state and emit light: we assume that when the phonon mode moves in the chaotic regime, the self-consistent Born approximation mechanism of the electron energy relaxation is inefficient.

When the atomic vibrations are in a chaotic state, the captured electrons are not relaxing their energy in the dot and the energy of the chaotic oscillator is being transferred to the rest of the vibrational modes of the environment of the dot, until the atomic vibration returns back to its laminar (or regular) motion.

We thus expect that the effect of the intermittence of the light emission of the individual quantum dots of the sample can be connected with the nonlinear properties of the atomic vibrations oscillators.

We illustrate the process of the transition in time, of the classical damped nonlinear oscillator driven by periodic force, to the state of the chaotic motion. This oscillator, given by the Newton's law (2) for the coordinate $y(t)$, starts at time $t = 0$ with the amplitude $y = 0$ and the derivative $y' = 0.1$. At the time of 300 the spectral density, defined as the absolute

$$y'' = -10y - y^3 - 0.1y^5 + 10.1\cos(t) - 0.0001y' \qquad (2)$$

value of the fourier transformation of *y*, corresponds to a regular motion with discrete frequencies, as it is seen in Fig. 2. At the time of 3000 the motion is already chaotic, as indicated by the presence of an interval of the frequency variable in which the fourier picture is a nonzero and continuous function (see Fig. 3). In the real process of blinking quantum dots the force, given by the 4th term on the right hand side of eq. (2), is not a simply periodic function. The present example only illustrates the process of the transition of the oscillator to the chaotic state by an applied force. Further modifications of the present simple model of the nonlinear oscillator are desirable in order to represent well the real systems.

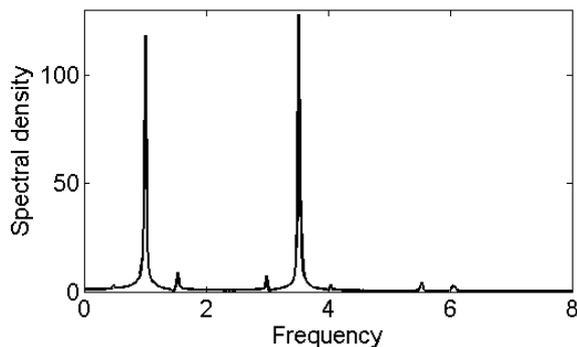

**Fig. 2** *Fourier picture of the time dependent coordinate of the nonlinear oscillator at time parameter equal to 300.*

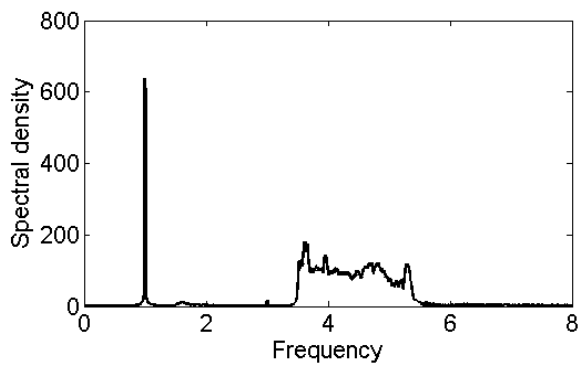

**Fig. 3** *Fourier picture of the time dependent coordinate of the nonlinear oscillator at time 3000.*

### 4. Conclusions

The purpose of the presentation is to turn the attention to the electron-phonon interaction in the zero-dimensional quantum nanostructures. The physical effects which we have mentioned are not to be confined only to the nanoparticles of inorganic materials. Similar effects are to be expected also in individual molecules, including the DNA molecules. Besides the problem of the energy balance analysis [23] in the problem of the electron energy relaxation in the zero-dimensional quantum nanostructures also the phenomena connected with the nonlinear oscillators of the atomic vibrations in the quasi-zer-dimensional nanostructures may need to be treated by the quantum kinetics analysis.

### Acknowledgements

This work was supported by the projects OC10007 of MŠMT and AVOZ10100520.